\documentclass[preprint,amsmath,amssymb,prb,showpacs]{revtex4}

\usepackage{graphicx,amssymb}
\usepackage{dcolumn}
\usepackage{bm}

\newcommand{\bfr}{{\bf r}}
\newcommand{\beq}{\begin{equation}}
\newcommand{\eeq}{\end{equation}}
\newcommand{\bea}{\begin{eqnarray}}
\newcommand{\eea}{\end{eqnarray}}

\def\bfr{{\bf r}}

\def\bfq{{\bf q}}
\def\bfp{{\bf p}}
\def\bfs{{\bf s}}

\begin{document}


\title{Exact results for a charged, harmonically trapped quantum gas
at arbitrary temperature and magnetic field strength}
\author{Patrick Shea and Brandon.~P. van Zyl}
\affiliation{Department of Physics, St.~Francis Xavier University, Antigonish, Nova Scotia, Canada B2G 2W5}


\begin{abstract}
An analytical expression for the first-order
density matrix of a charged, two-dimensional, harmonically 
confined quantum gas, 
in the presence of a constant magnetic field is derived.  In contrast
to previous results available in the literature, our
expressions are exact for any temperature and magnetic field strength.
We also present a novel factorization of the Bloch density matrix
in the form of a simple product with a clean separation of the
zero-field and field-dependent parts.   This factorization
provides an alternative way of analytically
investigating the effects of the magnetic field on the system,
and also permits the extension of our
analysis to other dimensions, and/or anisotropic confinement.
\end{abstract}


\maketitle

\section{Introduction}
Theoretical investigations of harmonically trapped ideal Fermi gases
have seen a renewed interest in recent years owing to the remarkable 
experimental advances made in the area of trapped, ultracold atoms. 
\cite{demarco}
Indeed, theorists now have an experimental realization of what is close to
being an ideal, inhomogeneous, quantum many-body Fermi system.  
Sophisticated magneto-optical traps now allow for
the possibility of ``tuning'' the dimensionality of these gases from three 
dimensions (3D) to quasi-2D or quasi-1D.  Thus, studying the properties of
essentially ideal, lower-dimensional many-body Fermi systems is now firmly in 
the realm of experimental fact, and not simply a matter of academic interest.
Furthermore, analytical results for these systems can be of great use in
the density-functional theory (DFT) of inhomogeneous Fermi systems, whereby
one can bypass the numerically expensive one-particle Schr\"odinger
equations.\cite{dft}

The ideal charged Bose gas (CBG) is the Bose analog of a charged Fermi 
system.  This model consists of a gas of spinless, charged bosons, 
coupled to an external, homogeneous magnetic field, and was first
investigated in 3D by Osborne,~\cite{osborne} Kosevitch,~\cite{kosevitch}
 and later by Schafroth.\cite{schafroth}
The uniform 2D CBG has also
been analytically studied quite extensively in the literature in light of 
its possible
connection to the theory of high-$T_c$ in the 
cuprates.\cite{may1,may2,frankel,tanatar,vanzyl_hutch,foulon}
To date, no detailed analytical analysis has been performed for 
the inhomogeneous case.
Since the confined 2D CBG is no longer forbidden
from undergoing a Bose-Einstein condensation (BEC) transition at 
low temperatures
(i.e., the Bogoliubov $1/k^2$ theorem is no longer 
applicable),~\cite{2Dvanzyl} an exact
analytical investigation of the thermodynamic and magnetic properties
(e.g., the Meissner-Ochsenfield (M-O) effect) of the inhomogeneous 
system would be of great interest.

The fundamental quantity from which the thermodynamic and magnetic properties
of the ideal quantum gases are derived is the first-order density matrix (FDM),
$\rho(\bfr,\bfr')$.  However, it
is highly non-trivial to obtain an exact expression for the FDM
(even at zero temperature) for all but the simplest of cases, viz., the
homogeneous ideal charged quantum gas.  
The introduction of a magnetic field
further complicates the problem, and it is only relatively 
recently that an exact analytical
expression for the zero temperature FDM of a {\em uniform} Fermi system 
coupled to a homogeneous magnetic field has become available.\cite{das}
Extensions 
of these results (i.e., to include the case of the CBG and finite
temperatures), have only been given in the last few years.\cite{vanzyl_hutch}
For the inhomogeneous ideal charged quantum gas, even the field-free
case at zero temperature is difficult.  Indeed, exact results 
for non-uniform systems at zero~\cite{brack_vanzyl,howard,murthy,akdeniz}
 and finite temperature~\cite{vanzyl_bhaduri,vanzyl_solo,wang}
are limited to the case of harmonic confinement.
To our knowledge, closed form, exact results for $\rho(\bfr,\bfr')$
for an ideal charged Fermi or Bose gas under general confinement, 
finite temperatures, and
arbitrary magnetic field strength, are not available.

The purpose of the present work is to help fill in this gap by
providing an analytical
expression for $\rho(\bfr,\bfr')$, generalized to treat {\em exactly}
the presence of a uniform external magnetic field and confining potential.
Our focus will be on
providing results for the 2D harmonically
confined quantum gas, although 
the general approach of our analysis does
allow for an extension to other dimensions, and anisotropic traps, 
should the need arise.\cite{note3}
The exact results of this paper should prove 
useful in the areas of current-density-functional theory,~
(CDFT)~\cite{vignale1,vignale2}
which is a rigorous
extension of DFT to inhomogeneous systems immersed in an
external magnetic field, and for the analytical investigation of the
magnetic and thermodynamic
properties of the CBG in the case of nonuniform systems.

The rest of our paper is organized as follows.  In the next section, 
we introduce the central theoretical tool used in our analysis, viz., 
the Bloch density matrix (BDM).  
In, Section III, we provide a 
derivation of the inverse Laplace transform of the BDM, which 
leads directly to the exact FDM for a Fermi or
Bose gas at
{\em any finite temperature and  magnetic field strength}.  
Section IV summarizes our main results and
offers a discussion of
how they may be applied in the context of CDFT and the
inhomogeneous 2D CBG.

\section{The Bloch Density Matrix}

The central theoretical tool used in our analysis is the zero temperature
BDM, $C_0(\bfr,\bfr';\beta)$,
which is related to the FDM through an inverse Laplace 
transform.\cite{semiclassical}
One of the key reasons for working with the BDM is that one does not
require explicit knowledge of the one-particle wave functions of the associated
trapping potential.  In addition, the zero temperature
BDM is independent of the quantum
statistics of the system, thereby allowing for an extremely robust
approach for treating either the Fermi or Bose gas.
Since a detailed discussion of the
BDM has already been given in our previous work,  we will only
present here the essential formalism required for a self-contained
statement of the problem, and refer the reader to 
Refs.~\onlinecite{vanzyl_hutch,brack_vanzyl,vanzyl_bhaduri,vanzyl_solo} for additional
details.

The zero temperature BDM is defined by

\begin{equation}
C_0(\bfr,\bfr';\beta) = \sum_{{\rm all i}} \psi_i^{\star}(\bfr')
\psi_i(\bfr)\exp(-\beta\epsilon_i)~,
\end{equation}
where the $\psi_i$'s and the one-particle energies $\epsilon_i$ are solutions
of the Schr\"odinger equation.  The constant $\beta$ above
is to be interpreted as a mathematical variable which in general is taken
to be complex, and not the inverse temperature $1/k_BT$.
The BDM satisfies the so-called Bloch equation

\begin{equation}
H_rC_0(\bfr,\bfr';\beta) = -\frac{\partial C_0(\bfr,\bfr';\beta)}
{\partial \beta}~,
\end{equation}
subject to the initial condition
\begin{equation}
C_0(\bfr,\bfr';0) = \delta(\bfr-\bfr')~.
\end{equation}
In this paper, 
\begin{equation}
H_r = \frac{(\bfp - e{\bf A}/c)^2}{2m} + \frac{1}{2}k(x^2+y^2)~,
\end{equation}
is the specific Hamiltonian we work with, where the magnetic field 
${\bf B}=\nabla\times{\bf A}$, 
is applied along the $z$-axis, and 
\begin{equation}
{\bf A} = \left(-\frac{1}{2}By,\frac{1}{2}Bx,0\right)~.
\end{equation}
Note that while $C_0(\bfr,\bfr';\beta)$ is gauge dependent, any physical
observable is necessarily gauge invariant.
By choosing a general functional form for $C_0(\bfr,\bfr';\beta)$,
the solution to Eqs.~(2-3), with the Hamiltonian (4), can be obtained
without having to specify the single-particle wave functions
or energies.  Such a solution has already 
been obtained by March and Tosi,~\cite{march_tosi}
which we now present in a more explicit form:
\begin{eqnarray}
C_0(\bfr,\bfr';\beta) &=&
\frac{m\omega_{\rm eff}}{2\pi\hbar}\frac{1}{\sinh(\hbar\omega_{\rm eff}\beta)}
e^{-\frac{im\omega_{\rm eff}}{\hbar}\frac{\sinh(\hbar\omega_c\beta)}{
\sinh(\hbar\omega_{\rm eff}\beta)}(xy'-yx')}\nonumber \\ 
&\times& e^{-[(x-x')^2+(y-y')^2]\frac{m\omega_{\rm eff}}{4\hbar}
\left[\coth(\hbar\omega_{\rm eff}\beta) + \frac{\cosh(\hbar\omega_c\beta)}
{\sinh(\hbar\omega_{\rm eff}\beta)}\right]}\nonumber  \\ 
&\times& 
e^{-[(x+x')^2+(y+y')^2]\frac{m\omega_{\rm eff}}{4\hbar}
\left[\coth(\hbar\omega_{\rm eff}\beta) - \frac{\cosh(\hbar\omega_c\beta)}
{\sinh(\hbar\omega_{\rm eff}\beta)}\right]}~,
\end{eqnarray}
where
\begin{equation}
\omega_c=\frac{eB}{2mc}~,~~~~~~\omega_{0} = \sqrt{\frac{k}{m}}~,~~~~~~
\omega_{\rm eff} = \sqrt{\omega_0^2 + \omega_c^2}~.
\end{equation}
Introducing the center-of-mass and relative
coordinates $\bfq$ and $\bfs$, respectively, 
\begin{equation}
\bfq = \frac{\bfr + \bfr'}{2}~,~~~~~~\bfs = \bfr-\bfr'~,
\end{equation}
allows us to write the BDM as
\begin{eqnarray}
C_0(\bfq,\bfs;\beta) &=& \frac{m\omega_{\rm eff}}{2\pi\hbar}
\frac{1}{\sinh(\hbar\omega_{\rm eff}\beta)}
e^{ -i\frac{m\omega_{\rm eff}}{\hbar}(q_ys_x-q_xs_y)
\frac{\sinh(\hbar\omega_c\beta)}{\sinh(\hbar\omega_{\rm eff}\beta)}}\nonumber
\\
&\times& e^
{-\frac{m\omega_{\rm eff}}{\hbar}\left[
q^2\left(\coth(\hbar\omega_{\rm eff}\beta) - \frac{\cosh(\hbar\omega_c\beta)}
{\sinh(\hbar\omega_{\rm eff}\beta)}\right)\right]}\nonumber \\
&\times& e^
{-\frac{m\omega_{\rm eff}}{\hbar}\left[
\frac{s^2}{4}\left(\coth(\hbar\omega_{\rm eff}\beta) + \frac{\cosh(\hbar\omega_c\beta)}
{\sinh(\hbar\omega_{\rm eff}\beta)}\right)\right]}~.
\end{eqnarray}
For later convenience, we now introduce the following definitions:
\begin{eqnarray}
A&=& q^2+\frac{s^2}{4}~,~~~~~B = \frac{1}{2}\left(\frac{s^2}{4}-q^2\right)
-\frac{i}{2}(q_ys_x - q_xs_y)~,~~~~~\omega = \frac{\omega_c}{\omega_{\rm eff}}~,
\end{eqnarray}
and scale all lengths and energies by 
$\ell_{\rm eff} = \sqrt{\hbar/m\omega_{\rm eff}}$ and
$\hbar\omega_{\rm eff}$, respectively.  The zero temperature BDM can
then be written in the more compact form
\begin{eqnarray}
C_0(\bfq,\bfs;\beta) &=&
\frac{1}{2\pi\sinh(\beta)} \exp\left(-A\coth(\beta) - B\frac{e^{-\omega\beta}}
{\sinh(\beta)} - B^{\star}\frac{e^{\omega\beta}}{\sinh(\beta)}\right)~,
\end{eqnarray}
where $B^{\star}$ denotes complex conjugation.  Equation (11) serves
as the starting point for the rest of our study, but it is worthwhile
pointing out that a novel factorization of the BDM can be performed.

First, let us re-write the BDM as follows
\begin{eqnarray}
C_0(\bfq,\bfs;\beta) &=&
\frac{1}{2\pi\sinh(\beta)} \exp
\left\{ -\left(q^2 + \frac{s^2}{4}\right)\coth(\beta) + 
\left(q^2-\frac{s^2}{4}\right)\frac{\coth(\omega\beta)}{\sinh(\beta)}
\right.\nonumber \\ 
&+&\left. i(q_ys_x-q_xs_y)\frac{\sinh(\omega\beta)}{\sinh(\beta)}\right\}~.
\end{eqnarray}
Making use of the trigonometric identities
\begin{eqnarray}
\frac{\cosh(\beta)-\cosh(\omega\beta)}{\sinh(\beta)}
&=& \tanh(\beta/2)-2\frac{\sinh^2(\omega\beta/2)}{\sinh(\beta)}~,\nonumber \\
\frac{\cosh(\beta)+\cosh(\omega\beta)}{\sinh(\beta)}
&=& \coth(\beta/2)+2\frac{\sinh^2(\omega\beta/2)}{\sinh(\beta)}~,
\end{eqnarray}
in Eq.~(12) gives 
\begin{eqnarray}
C_0(\bfq,\bfs;\beta) &=&
\frac{1}{2\pi\sinh(\beta)}
e^{\left\{
-q^2\tanh(\beta/2) - \frac{s^2}{4}\coth(\beta/2)\right\}}
e^{U_c(\bfq,\bfs;\beta)}~,
\end{eqnarray}
where 
\begin{equation}
U_c(\bfq,\bfs;\beta) \equiv 2\left(q^2-\frac{s^2}{4}\right)
\frac{\sinh^2(\omega\beta/2)}{\sinh(\beta)} +
i\left( q_xs_x - q_xs_y\right)\frac{\sinh(\omega\beta)}{\sinh(\beta)}~.
\end{equation}
The ``effective potential'' $U_c(\bfq,\bfs;\beta)$ explicitly
includes all of the magnetic field
dependence, and there is a clean separation of the BDM into field free
and field dependent parts.  In particular, setting $\omega_c=0$ 
in Eq.~(15) immediately gives $U_c=0$ and the BDM, (14),  reduces
to that of a 2D harmonically trapped 
system.\cite{brack_vanzyl,vanzyl_bhaduri,vanzyl_solo}
This factorization is reminiscent
of the introduction of an effective potential in Ref.~\onlinecite{pfalzner},
which was motivated
by the desire to improve the Thomas-Fermi approximation to potentials
which are varying too rapidly in some regions of space.\cite{note4}
Viewing the magnetic field as an additional 1D confining potential suggests
a similar interpretation in the present context; that is, going beyond
$\omega_c=0$ may be achieved through the introduction of some effective
potential, $U_c$, which encodes the magnetic field dependence.  
Irrespective of this
suggestive connection however, 
Eq.~(14) here should be viewed as a more direct route to generalizing our
results below to other dimensions, and allowing for a more transparent
analytical investigation of the effects of the magnetic field in the 
weak/high field limits, along with anisotropic confinement, 
should this be desired.  

We are now in a position to see why the BDM provides such a 
universal scheme to investigate either the Fermi or Bose gases.
While the zero temperature BDM is independent of the quantum statistics, 
at finite temperature, the BDM for the Fermi system is obtained via
($k_B=1$)~\cite{semiclassical}
\begin{equation}
C_T(\bfq,\bfs;\beta) = C_0(\bfq,\bfs;\beta)\frac{\pi\beta T}{\sin(\pi\beta T)}
~,~~~({\rm fermions})~,~
\end{equation}
whereas for bosons, it is given by
\begin{equation}
C_T(\bfq,\bfs;\beta) = C_0(\bfq,\bfs;\beta)\frac{-\pi\beta T}{\tan(\pi\beta T)}
~,~~~({\rm bosons})~.
\end{equation}
Therefore, aside from the different temperature dependent factors in Eqs.~(16,17),
it is clear that only the $T=0$ BDM is required to study either quantum gas.
\section{The First-Order Density Matrix}
\subsection{Fermi gas}
The (spin-averaged) FDM at finite temperature is obtained by a two-sided
inverse Laplace transform of the finite temperature BDM.
The inverse
Laplace transform must be two-sided to allow for the dual variable 
to go negative. 
Specifically, we have~\cite{note2}
\begin{equation}
\rho(\bfq,\bfs;T) = \mathfrak{L}^{-1}_{\mu}\left[\frac{2}{\beta}C_T(\bfq,\bfs;\beta)\right]~,
\end{equation}
where $\mu$ is the chemical potential, which at fixed
$\omega_c$, is determined by particle number conservation.
As we have discussed before,~\cite{brack_vanzyl,vanzyl_bhaduri,vanzyl_solo}
it is very difficult to perform
the inverse Laplace transform by simply substituting the finite temperature
BDM, as given by Eqs.~(11,16), into Eq.~(18).  
In order to proceed any further analytically,
one requires the following identities
\begin{equation}
\exp(-A\coth(\beta)) = \sum_{k=0}^{\infty} L_k(2A)e^{-A}\left\{
e^{-2k\beta} - e^{-2(k+1)\beta}\right\}
\end{equation}
\begin{equation}
\exp\left(-\frac{Be^{-\omega\beta}}{\sinh(\beta)}\right) =
\sum_{m=0}^{\infty} \sum_{i=0}^{m}
\frac{ (-2B e^{-(\omega-1)\beta})^i}{i!}
\left(\begin{array}{c}
m\\ m-i
\end{array}\right)\left\{e^{-2m\beta} - e^{-2(m+1)\beta}\right\}
\end{equation}
\begin{equation}
\exp\left(-\frac{B^{\star}e^{\omega\beta}}{\sinh(\beta)}\right) =
\sum_{n=0}^{\infty} \sum_{j=0}^{n}
\frac{ (-2B^{\star} e^{(\omega+1)\beta})^j}{j!}
\left(\begin{array}{c}
n\\ n-j
\end{array}\right)\left\{e^{-2n\beta} - e^{-2(n+1)\beta}\right\}~,
\end{equation}
where $L_{l}(x)$ is a Laguerre polynomial.
Utilizing these identities in Eq.~(11) gives
\begin{eqnarray}
C_0(\bfq,\bfs;\beta) &=& \frac{1}{\sinh(\beta)}
\sum_{l=0}^{\infty}\sum_{m=0}^{\infty}\sum_{n=0}^{\infty}
\sum_{i=0}^{m}\sum_{j=0}^{n}
L_{l}(2A)e^{-A}\frac{(-2B)^i}{i!}\frac{(-2B^{\star})^j}{j!}
\left(
\begin{array}{c}
m\\m-i\end{array}\right)
\left(
\begin{array}{c}
n\\n-j\end{array}\right) \nonumber \\
&\times&
\left\{
e^{(-2l -2m - 2n +i+j)\beta+(j-i)\omega\beta} - 
3 e^{(-2l-2m-2n+i+j-2)\beta +(j-i)\omega\beta}
\right\}\nonumber \\
&\times&
\left\{
3e^{(-2l -2m - 2n +i+j-4)\beta+(j-i)\omega\beta} -  
e^{(-2l-2m-2n+i+j-6)\beta+(j-i)\omega\beta}
\right\}~.
\end{eqnarray}
Notice that all of the $\beta$ dependence is now contained in the exponential 
factors and the (two-sided) inverse Laplace transform
is now tractable.  The mathematical details of this transform closely follows
our earlier work,~\cite{vanzyl_hutch,vanzyl_bhaduri,vanzyl_solo}
so here we will simply give the final result, namely,
\begin{eqnarray}
\rho(\bfq,\bfs;T) &=&
\frac{2}{\pi}\sum_{l=0}^{\infty}\sum_{m=0}^{\infty}\sum_{n=0}^{\infty}
\sum_{i=0}^{m}\sum_{j=0}^{n}\sum_{k}
L_{l}(2A)e^{-A}\frac{(-2B)^i}{i!}\frac{(-2B^{\star})^j}{j!}
\left(
\begin{array}{c}
m\\m-i\end{array}\right)
\left(
\begin{array}{c}
n\\n-j\end{array}\right) \nonumber \\
&\times&
\mathfrak{F}_k(l,m,n,i,j)~,
\end{eqnarray}
where all of the temperature dependence is encoded in the Fermi-like
function
\begin{equation}
\mathfrak{F}_k(l,m,n,i,j) \equiv
\frac{1}{\exp\left(\frac{k+2(l+m+n)-i-j-(j-i)\omega-\mu}{T}\right) + 1}~~~~~
({\rm fermions})~,
\end{equation}
and the $k$-summation is over $k= 1, 3, 5$.
Equation (23) is the central result of this paper and gives the
exact FDM for an ideal, harmonically trapped 2D Fermi gas at arbitrary 
temperature and magnetic field strength.  
Putting $\bfs = 0$ immediately yields the
spatial density of the system.  
\begin{figure}[ht]
\scalebox{0.7}{\includegraphics*{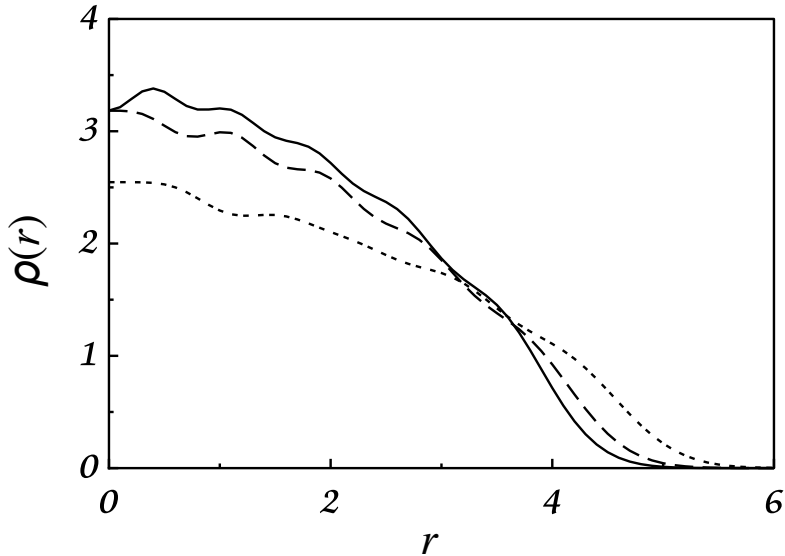}}
\caption{\label{fig1}  Plot of the spatial density for $N=110$
fermions at zero-temperature and various magnetic field strengths.
The solid curve is for $\omega=0$, the dashed curve for $\omega = 0.45$
and the dotted curve for $\omega = 0.70$.  All lengths and energies have
been scaled as discussed in the text}
\end{figure}
It is certainly worthwhile re-emphasizing that {\em all} previous results 
pertaining to this system can now be obtained from (23)
upon taking various limits.  For example, in the uniform case at zero
temperature, Eq.~(23) can be shown to reduce to (with dimensional constants
recovered)~\cite{das,vanzyl_hutch}
\begin{equation}
\rho\left(\bfq+\frac{\bfs}{2},\bfq - \frac{\bfs}{2}\right) =
\frac{2m\omega_c}{\pi\hbar}
e^{-i(m\omega_c/\hbar)(x'y-y'x)}
e^{-(m\omega_c/2\hbar)s^2}
L^{(1)}_{n_F-1}\left(\frac{m\omega_c}{\hbar}s^2\right)~.
\end{equation}
As an illustrative numerical example, we present in Fig.~1, the 
spatial density for
$N=110$ particles at zero-temperature and various magnetic field strengths.
It is important to note that while the summations at finite temperature
in Eq.~(23) look somewhat formidable, any practical numerical implementation
requires only a relatively small number of terms. 
Figure 1, for example,
required only a few minutes to plot using a nominally equipped
PC running a generic flavor of Unix.
Consequently, finite temperature effects are readily studied, should the need
arise.\cite{note1}  Note also that the relative ease for
which we were able to write down $\rho(\bfq,\bfs;T)$ should not be used to 
conclude that the calculation is trivial.  In particular, 
it is notoriously difficult
to treat finite temperature effects exactly in the Fermi gas 
owing to the fact that
one cannot express the Fermi distribution function as a convergent power 
series, except at very high temperatures.\cite{wang}  Rather, one should view
our almost immediate statement of the FDM as a testament to the
utility of the inverse Laplace transform technique.
\subsection{Bose gas}
The power of the inverse Laplace transform technique is also
apparent if one wishes to extend our results to the case of a harmonically
confined 2D CBG.  Indeed, one can immediately write down the final 
expression for
the FDM, with the
only changes being a change in the sign in front of unity in the 
denominator of Eq.~(24), viz.,
\begin{equation}
\mathfrak{F}_k(l,m,n,i,j) \equiv
\frac{1}{\exp\left(\frac{k+2(l+m+n)-i-j-(j-i)\omega-\mu}{T}\right) - 1}~~~~~
({\rm bosons})~,
\end{equation}
and the elimination of the factor of two in Eq.~(18) (i.e., the bosons are 
spinless).
Thus, with no additional work, we also have an exact, closed form 
expression for the FDM of the trapped
2D CBG at arbitrary temperature and magnetic field strength.  
Of course, setting $\omega_0=0$ reproduces the recently
obtained finite temperature results corresponding to the uniform 2D 
CBG found in Ref.~\onlinecite{vanzyl_hutch}.
\section{Summary and Future Work}
We have derived an analytical expression for the FDM
of an ideal, harmonically trapped charged 2D Fermi or Bose gas at finite
temperature and arbitrary magnetic field strength. 
To our knowledge,
this is the only example where such an exact expression has been
obtained for an inhomogeneous quantum gas.   Aside from their inherent 
technical merit, we believe that our results now open up several other 
fruitful avenues of investigation, which are outside the intended
scope of this paper.   

First it would be interesting to investigate the properties of the 2D
Thomas-Fermi kinetic energy functional for the case of finite magnetic field.
The motivating factor behind this suggestion lies in the remarkable fact
that for zero-field, the 2D Thomas-Fermi kinetic energy
functional leads to the {\em exact} quantum mechanical kinetic energy
(i.e. without gradient corrections) when integrated over all 
space.  This
non-trivial result was only recently discovered by 
Brack and one of us.\cite{brack_vanzyl}
From Eq.~(25), it can be shown that the local-density approximation 
(LDA)~\cite{dft} of the magnetic 2D kinetic
energy functional is identical in form to the zero-field case, but now
the magnetic field is encoded implicitly by the 
density.\cite{das,vanzyl_hutch}
Thus, determining whether the 2D magnetic-Thomas-Fermi kinetic energy 
functional is also exact 
(i.e., similar to its zero-field counterpart without gradient corrections)
would be very interesting.
It would also be illustrative to study the
2D exchange energy (i.e., suitable for the study of parabolically confined
quantum dots in a magnetic field) via the exact FDM and compare the 
integrated, and spatial properties against the commonly used
LDA of CDFT.  This type of
comparison has already been undertaken for the zero magnetic field 
case~,\cite{vanzyl_solo}
and would be an equally worthwhile endeavor for the finite-field case.
Furthermore, having an exact expression for $\rho(\bfr,\bfr';T)$ also allows
for the perturbative study of the effects of particle-particle interactions, 
similar to what has already been
performed for the $\omega_c=0$ case in Ref.~\onlinecite{vanzyl_bhaduri}.

As for the CBG, the most obvious application of
our results will be in investigating the thermodynamic and magnetic
properties of the inhomogeneous system.  In contrast to the uniform 2D CBG,
the trapping potential stabilizes the system to density and phase 
fluctuations and
allows for the possibility of a transition to a BEC.\cite{2Dvanzyl}
We recall here that the uniform 2D CBG does exhibit an essentially 
perfect M-O effect, in spite of the {\em absence} of a 
BEC state.\cite{vanzyl_hutch}
Thus, analytically studying the connection
between the onset of the M-O effect and the BEC phase in the trapped
system is, in our opinion, an important problem.
\begin{acknowledgments}
This research was supported in part by a  Discovery grant 
from the National Sciences and Engineering
Research Council (NSERC) of Canada.  Patrick Shea would also like to acknowledge
financial support from an NSERC undergraduate student research award (USRA).
\end{acknowledgments}


\begin{thebibliography}{90}
\bibitem{note3}
Our restriction to 2D is for the following reasons: 
(i) the 2D calculation is the most
transparent application of the inverse Laplace transform technique and (ii)
we expect that the most relevant applications of our work will be to the
inhomogeneous 2D Fermi and charged Bose gases.
\bibitem{note2}
At zero temperature, Eq.~(18) still holds, but with $C_0$ instead of
$C_T$, and the inverse Laplace transform is now one-sided.
\bibitem{note1}
Of course, at high temperatures, quantum effects (i.e., shell effects) are
already washed out, and it makes no sense to continue using Eq.~(23).  
See e.g., Ref.~\onlinecite{vanzyl_bhaduri} for an illustrative example 
of this scenario.
\end{thebibliography}
\end{document}